\documentclass[twocolumn,aps,prl]{revtex4}
\usepackage{amsmath}
\usepackage{graphicx}

\begin{document}

\title{Heat Transfer between Graphene and Amorphous ${\rm SiO_2}$}
\author{ B.N.J. Persson$^{1,2}$ and H. Ueba$^{1}$}

\affiliation{$^1$ Division of Nanotechnology and New Functional Material Science,
Graduate School of Science and Engineering,
University of Toyama, Toyama, Japan}
\affiliation{$^2$IFF, FZ-J\"ulich, 52425 J\"ulich, Germany, EU}

\begin{abstract}
We study the heat transfer between graphene and amorphous ${\rm SiO_2}$. 
We include both the heat transfer from the area of
real contact, and between the surfaces in the
non-contact region. We consider the radiative heat transfer associated with the
evanescent electromagnetic waves which exist outside of all bodies, and the heat transfer
by the gas in the non-contact region. We find that the dominant contribution to the
heat transfer result from the area of real contact, and the calculated value of
the heat transfer coefficient is in good agreement 
with the value deduced from experimental data.
\end{abstract}

\maketitle

\pagestyle{empty}

%%%%%%%%%%%%%% main text %%%%%%%%%%%%%%%%
%\begin{multicols}{2}

Graphene, the recently isolated 2-dimensional (2D) carbon material with unique properties due to its linear electronic
dispersion, is being actively explored for electronic applications\cite{Geim}. 
Important properties are the high mobilities reported especially in suspended graphene, the fact that graphene is the ultimately thin
material, the stability of the carbon-carbon bond in graphene, the ability to induce a bandgap by electron confinement in graphene
nanoribbons, and its planar nature, which allows established pattering and etching techniques to be applied.

Recently it has been found that the heat generation in graphene field-effect transistors can result 
in high temperature and
device failure\cite{Freitag}. Thus, it is important to understand the the mechanisms which influence the heat flow. 
Because of surface roughness 
the graphene will only make partial contact with the  ${\rm SiO_2}$ substrate,
which will reduce the heat transfer coefficient as compared to the perfect contact case.
In Ref. \cite{Freitag} the temperature profile in the graphene under current was obtained 
by studying a Raman active phonon band of graphene (the position of the phonon band is strongly dependent on the local temperature
and can be used as a microscopic thermometer). The heat transfer coefficient between graphene 
and the ${\rm SiO_2}$ substrate was determined by modeling the heat flow using the standard
heat flow equation with the heat transfer coefficient as the only unknown quantity. The authors found that using
a constant (temperature independent) heat transfer coefficient  $\alpha \approx 2.5\times 10^7 \ {\rm W/m^2K}$
resulted in calculated temperature profiles in the graphene in good agreement with experiment.

The heat transfer coefficient between graphene and a perfectly flat ${\rm SiO_2}$ substrate has not been measured
directly, but measurements of the heat transfer between carbon nanotubes and sapphire 
by Maune et al\cite{Maune} indicate that it may be of order $\alpha \approx 8\times 10^8  \ {\rm W/m^2K}$.
This value was deduced indirectly by measuring the breakdown voltage of carbon nanotubes, which could be related to the
temperature increase in the nanotubes. Molecular dynamics calculations\cite{Pop} for nanotubes on amorphous ${\rm SiO_2}$  gives 
$\alpha \approx 3 \times 10^8  \ {\rm W/m^2K} $, where we have assumed the contact width between the nanotube and
the substrate to be $1/5$ of the diameter of the nanotube. A similar value can be deduced from theory\cite{PerssonTBP}.

Here we
present a theoretical study of the different mechanisms which determine the heat transfer between  
graphene and an amorphous ${\rm SiO_2}$ substrate. 
We study the contribution to the heat transfer not just from the area of real contact,
but also the heat transfer across the non-contact surface area, in particular the contribution from the fluctuating
electromagnetic field, which surrounds all solid objects\cite{rev1}, and heat transfer via the surrounding gas. 

\begin{figure}[tbp]
\includegraphics[width=0.4\textwidth,angle=0]{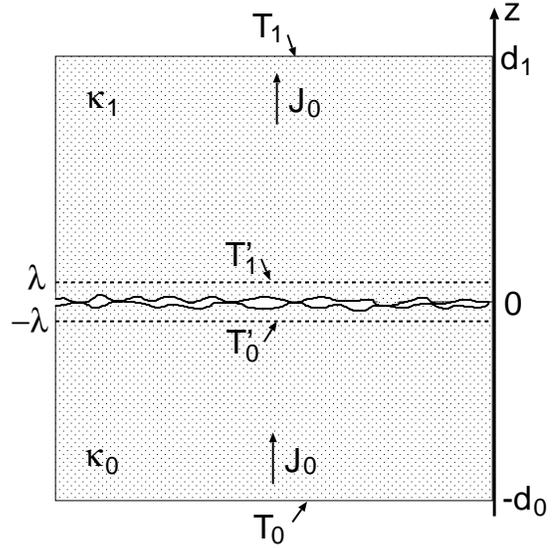}
\caption{
Two elastic solids with nominally flat surfaces squeezed together with the nominal
pressure $p_0$.
The heat current $J_{\rm z}({\bf x})$ at the contacting interface varies strongly with
the coordinate ${\bf x} = (x,y)$ in the $xy$-plane. The average heat current is denoted
by $J_0= \langle J_{\rm z}({\bf x})\rangle$.}
\label{contactblock}
\end{figure}

Consider two elastic solids (rectangular blocks) with randomly rough surfaces
squeezed in contact as illustrated in Fig. \ref{contactblock}.
Assume that the temperature at the outer surfaces $z=-d_0$ and $z=d_1$
is kept fixed at $T_0$ and $T_1$, respectively, with $T_0 > T_1$.
Close to the interface
the heat current will vary rapidly in space, ${\bf J} = {\bf J} ({\bf x},z)$, where ${\bf x}=(x,y)$ denotes the
lateral coordinate in the $xy$-plane. Far from the interface we will assume that the heat current is constant
and in the $z$-direction, i.e., ${\bf J}= J_0\hat z$.
We denote the average distance between the macro asperity contact regions by $\lambda$ (see Ref. \cite{PSSR}). We assume that
$\lambda << L$, where $L$ is the linear size of the apparent contact
between the elastic blocks. The temperature at a distance $\sim \lambda$ from the contacting interface will
be approximately independent of the lateral coordinate ${\bf x} = (x,y)$ and we denote this temperature by
$T_0'$ and $T_1'$ for $z= -\lambda$ and $z=\lambda$, respectively. The heat current for $|z| >> \lambda$
is independent of ${\bf x}$ and can be written as (to zero order in $\lambda /d_0$ and $\lambda /d_1$):
$$J_0=-\kappa_0 {T_0'-T_0\over d_0} = -\kappa_1 {T_1-T_1'\over d_1},\eqno(1)$$
where $\kappa_0$ and $\kappa_1$ are the heat conductivities of the two solid blocks.
We assume that the heat transfer across the interface is proportional to
$T_0'-T_1'$ and we define the heat transfer coefficient $\alpha$ so that
$$J_0=\alpha (T_0'-T_1')\eqno(2)$$
Combining (1) and (2) gives
$$J_0={T_0-T_1\over d_0 \kappa_0^{-1} +d_1 \kappa_1^{-1}+\alpha^{-1}}\eqno(3)$$
This equation is valid as long as $\lambda << L$ and $\lambda << d_0, \ d_1$.
In our application the upper solid is graphene which is only one atom thick. But because of the
very high in-plane thermal conductivity of graphene compared to the substrate (${\rm SiO_2}$)
the temperature in the graphene will be nearly constant, and we can define the heat
transfer coefficient $\alpha$ using (2) in this case too.

\begin{figure}[tbp]
\includegraphics[width=0.5\textwidth,angle=0]{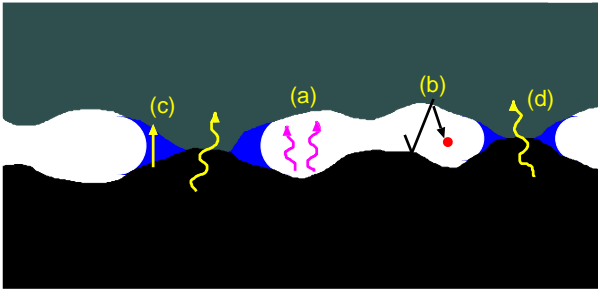}
\caption{
The heat transfer between two solids occur via (a) the electromagnetic field (e.g., photon tunneling),
(b) from heat diffusion or ballistic energy transfer in the surrounding gas (or liquid), via (c)
capillary bridges and (d) via the area of real contact.}
\label{heattransferProcesses}
\end{figure}

In Fig. \ref{heattransferProcesses} we show the different heat transfer processes we discuss below. 
We first study the radiative contribution (a) to the heat transfer\cite{PLV,rev1,Rotkin}.
The heat flux per unit area between two black-bodies separated by $d>> d_T= c\hbar /k_BT$ is given by
the Stefan-Boltzmann law
$$J_0 = {\pi^2 k_{\rm B}^4 \over 60 \hbar^3 c^2} \left (T_0^4-T_1^4\right )$$
where $T_0$ and $T_1$ are the temperatures of solids ${\bf 1}$ and ${\bf 2}$, respectively,
and $c$ the light velocity. In this limiting case the heat transfer between the bodies is determined by
the propagating electromagnetic waves radiated by the bodies and does not depend on the separation
$d$ between the bodies. Electromagnetic waves (or photons) always exist outside any body due to thermal
or quantum fluctuations of the current density inside the body. The electromagnetic field created by
the fluctuating current density exists also in the form of evanescent waves, which are damped exponentially
with the distance away from the surface of the body. For an isolated body, the evanescent waves do not give
a contribution to the energy radiation. However, for two solids separated by $d < d_{T}$, the heat transfer may
increase by many orders of magnitude due to the evanescent electromagnetic waves--this is often referred to
as photon tunneling.

For short separation between two solids with flat surfaces ($d << d_{T}$), the heat current due to the
evanescent electromagnetic waves is given by\cite{rev1}
$$J_0 = {4\over (2\pi)^3} \int_0^\infty d\omega \ \left (\Pi_0(\omega)-\Pi_1(\omega)\right )$$
$$\times \int d^2q \ e^{-2qd} {{\rm Im} R_0({\bf q},\omega) {\rm Im} R_1({\bf q},\omega) \over
|1-e^{-2qd} R_0({\bf q},\omega)R_1({\bf q},\omega) |^2}\eqno(4)$$
where
$$\Pi (\omega) = \hbar \omega \left (e^{\hbar \omega /k_{\rm B}T}-1\right )^{-1}$$
and the reflection factor
$$R({\bf q},\omega) = {\epsilon ({\bf q},\omega) -1 \over \epsilon ({\bf q},\omega) + 1}$$
where $\epsilon ({\bf q},\omega)$ is the dielectric function.
From (4)  it follows that if $R_0({\bf q},\omega)$ and $R_1({\bf q},\omega)$
are independent of ${\bf q}$, the heat current scale as $1/d^2$ with the separation between the solid surfaces.

\begin{figure}[tbp]
\includegraphics[width=0.45\textwidth,angle=0]{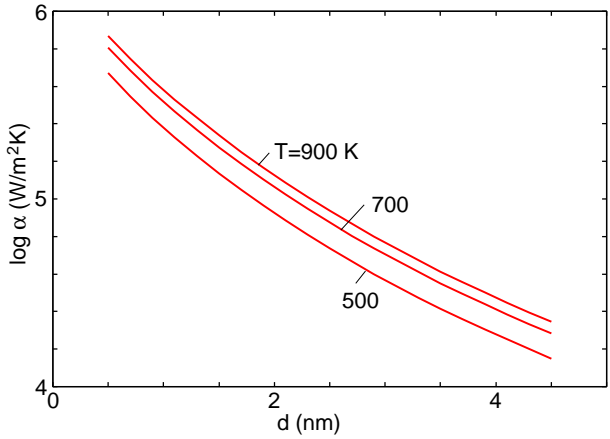}
\caption{
The logarithm of the heat transfer coefficient, between graphene and an 
amorphous ${\rm SiO_2}$ solid with a flat surface,
as a function of the separation (in nanometer) $d$ between the solids. For the temperatures $T=500$, $700$ and
$900  \ {\rm K}$.}
\label{distance.nm.alpha}
\end{figure}

We now apply (4) to graphene adsorbed on a nominally flat surface of amorphous ${\rm SiO_2}$.
The graphene dielectric function was recently calculated using the mean field approximation where the electric
field acting on an electron is the sum of the external electric field, and the induced field from
the other electrons\cite{HS}. In the calculation it was assumed that the one-particle energy eigenvalues are linearly
related to the wavevector ${\bf k}$ via $\epsilon_{s{\bf k}}=s \gamma |{\bf k}|$, where $s=\pm1$ indicate the 
conduction and valence bands, and where $\gamma \approx 6.5 \ {\rm eV \AA}$ is a band parameter. In this model 
the Fermi wavevector $k_{\rm F} = (\pi n)^{1/2}$, and the Fermi energy $E_{\rm F} = \gamma k_{\rm F}$,
where $n$ is the number of 2D carriers (electron or hole) per unit area.
Note that in the absence of doping (and for vanishing gate voltage) $n=0$. In the application below
the excitation energy $\hbar \omega$ is of order the thermal energy $k_{\rm B} T$ while the wavevector typically
is of order $1/d$, where $d$ is the average separation between the 
graphene and the substrate surface. Under these conditions
$ \hbar \omega << \gamma q$ which is equivalent to $q/k_{\rm F} >> \hbar \omega /E_{\rm F}$. In this limit 
and assuming we can neglect the influence of the temperature on the dielectric properties, for $q<2k_{\rm F}$ the
dielectric function takes the following simple form\cite{HS}:  
$$\epsilon({\bf q},\omega) = 1+ \left (\pi n \right )^{1/2} {4e^2\over \gamma q} \left (1+i{\hbar \omega \over \gamma q}  \left [1-\left ( {q\over 2k_{\rm F}}\right )^2 \right ]^{1/2}
\right ).$$
For  $q>2k_{\rm F}$ and $ \hbar \omega << \gamma q$ 
the imaginary part of $\epsilon$ vanish so there will be no contribution to the heat transfer
from  $q>2k_{\rm F}$ (note: the contribution from the pole of $R({\bf q},\omega)$ is negligible). 
In the numerical calculations presented below we assume $n=10^{13} \ {\rm cm}^{-2}$ giving $k_{\rm F} = 0.056 \ {\rm \AA}^{-1} $ 
and $E_{\rm F} = 0.36 \ {\rm eV}$.

The optical properties of (amorphous) silicon dioxide (SiO$_2$) can be described using an oscillator
model\cite{optical}
$$\epsilon (\omega) = \epsilon_\infty + {a\over \omega_a^2 -\omega^2 -i\omega \gamma_a}+{b\over \omega_b^2 -\omega^2 -i\omega \gamma_b}$$
The frequency dependent term in this expression is due to optical phonon's.
The values for the parameters $\epsilon_\infty$, $(a,\omega_a,\gamma_a)$
and $(b,\omega_b,\gamma_b)$ are given in Ref. \cite{optical}.

If surface roughness occur so that the separation $d$ varies
with the coordinate ${\bf x}=(x,y)$ we have $\alpha \approx \langle \alpha (d) \rangle$,
where $\langle .. \rangle$ stands for ensemble average, or average over the whole surface area.

In the preset case the heat transfer is associated with thermally excited optical (surface) phonon's
in  ${\rm SiO_2}$ and electron-hole pairs in the graphene. That is, the electric field of a
thermally excited optical phonon in  ${\rm SiO_2}$ excites an electron-hole pair in the graphene, leading to energy
transfer. The excitation transfer occurs in both directions but if one solid is hotter than the other,
there will be a net transfer of energy from the hotter to the colder solid. 

In Fig. \ref{distance.nm.alpha} we show the logarithm of the heat transfer coefficient, 
between graphene and an amorphous ${\rm SiO_2}$ solid with a flat surface,
as a function of the separation (in nanometer) $d$ between the solids. 
The results have been obtained from (4) for the temperatures $T=500$, $700$ and
$900  \ {\rm K}$. For $d \approx 1 \  {\rm nm}$ we obtain $\alpha \approx 3\times 10^5  \ {\rm W/m^2 K}$.
The heat transfer coefficient deduced from the experiment\cite{Freitag} is $\alpha_{\rm exp} \approx 2.5 \times 10^7 \ {\rm W/m^2 K}$,
and we conclude that the field coupling gives a negligible contribution to the heat transfer.
Note also that the photon tunneling contribution to $\alpha$
depends strongly on the temperature (it increases with a factor of $\approx 2$ as the
temperature increases from $400 \ {\rm K}$ to $900 \ {\rm K}$), while the experimental data\cite{Freitag} could be fit with a 
temperature independent $\alpha$.

Let us now consider the contribution (b) to the heat transfer from the surrounding gas.
Consider two solids with flat surfaces separated by a distance $d$. Assume that the solids are surrounded by a gas.
Let $\Lambda$ be the gas mean free path. If $d >> \Lambda$ the heat transfer between the solids occurs via
heat diffusion in the gas. If $d << \Lambda$ the heat transfer occurs by ballistic propagation of gas molecules
from one surface to the other. In this case gas molecules reflected from the hotter surface will have (on the average)
higher kinetic energy that the gas molecules reflected from the colder surface. This will result in heat transfer from
the hotter to the colder surface. The heat transfer coefficient is approximately given by\cite{Bahrami1}
$$\alpha  \approx {\kappa_{\rm gas} \over d + \Lambda}$$
For air (and most other gases) at the normal atmospheric pressure and at room temperature $\Lambda \approx 65 \ {\rm nm}$
and $\kappa_{\rm gas} \approx 0.02 \ {\rm W/mK}$.
For contacting surfaces with surface roughness we get 
$$\alpha \approx \kappa_{\rm gas} \langle (d+\Lambda )^{-1} \rangle\eqno(5)$$
where $\langle .. \rangle$ stand for ensemble average or averaging over the surface area.

In the present application the
surface separation is of order $\approx {\rm nm}$  so we can neglect the $d$-dependence in (5) and get
$\alpha \approx \kappa_{\rm gas}/\Lambda \approx 3\times 10^5 \ {\rm W/m^2 K}$, which is similar to the contribution
from the electromagnetic coupling but much smaller than the heat transfer coefficient deduced from the experiment\cite{Freitag}.
Note also that $\kappa_{\rm gas}$ (and hence $\alpha_{\rm gas}$) 
depends strongly on the temperature (it increases with a factor of $\approx 2$ as the
temperature increases from $300 \ {\rm K}$ to $600 \ {\rm K}$), while the experimental data\cite{Freitag} could be fit with a 
temperature independent $\alpha$.

We now study the contribution (c) to the heat transfer from capillary bridges.
If the solid walls are wet by water, in a humid atmosphere capillary bridges will form spontaneous
at the interface in the vicinity of the asperity contact regions. For very smooth and hydrophilic surfaces
the fluid (in this case water) may occupy a large region between the surfaces and will then
dominate the heat transfer between the solids. Similarly, contamination layers (mainly organic molecules)
which cover most natural surfaces may form capillary bridges between the contacting solids, and
contribute in an important way to the heat transfer coefficient. The fraction of the interfacial
surface area occupied by fluid bridges, and the separation between the solids in the fluid
covered region, can be calculated using the theory developed in Ref. \cite{PerssonCapillary}. From this one can calculate the
contribution to the heat transfer using:
$$\alpha \approx \kappa_{\rm liq} \langle d^{-1} \rangle$$
For the present system we do not expect capillary bridges to be important because the experiment was performed in
dry nitrogen atmosphere.

\begin{figure}[tbp]
\includegraphics[width=0.45\textwidth,angle=0]{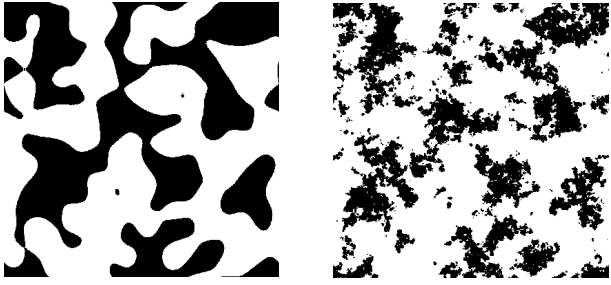}
\caption{
The contact region (black area) between two elastic solids observed at
low (left) and high (right) magnification. The contact resistance
depends mainly on the long-wavelength roughness, and can usually be calculated
accurately from the nature of the contact observed at low magnification (left).}
\label{HeatArea}
\end{figure}

The study above shows that the contribution to $\alpha$ from processes (a)-(c) are much smaller than the
observed heat transfer coefficient. Thus the heat transfer must be dominated by the only remaining process,
namely heat flow via the area of real contact. 
Recent contact mechanics studies have shown
that for elastic contact, the contact regions observed at atomic
resolution may be just a few atoms wide, i.e., the diameter of the
contact regions may be of the order of $\sim 1 \ {\rm
nm}$\cite{Chunyan,Hyun,Nature1}. The heat transfer via such small
junctions may be very different from the heat transfer through
macroscopic sized contact regions, where the heat transfer usually
is assumed to be proportional to the linear size of the contact
regions (this is also the prediction of the macroscopic heat
diffusion equation), rather than the contact area. In particular, if
the typical phonon wavelength involved in the heat transfer becomes
larger than the linear size of the contact regions (which will
always happen at low enough temperature) the effective heat transfer
may be strongly reduced. Similarly, if the phonons mean free path is
longer than the linear size of the contact regions, ballistic
(phonon) energy transfer may occur which cannot be described by the
macroscopic heat diffusion equation. However, as shown in Ref. \cite{PLV},
for macroscopic solids the
thermal (and electrical) contact resistance is usually very
insensitive to the nature of the contact regions observed at the
highest magnification, corresponding to atomistic (or nanoscale)
length scales. In fact, the heat transfer is determined mainly by
the nature of the contact regions observed at lower magnification
where the contact regions appear larger (see Ref. \cite{PLV} and Fig. \ref{HeatArea}). 
Thus, the thermal contact resistance of
macroscopic solids usually does not depend on whether the heat
transfer occur by diffusive or ballistic phonon propagation, but
rather the contact resistance is usually determined mainly by the
nature of the contact regions observed at relative low
magnification.

\begin{figure}[tbp]
\includegraphics[width=0.5\textwidth,angle=0]{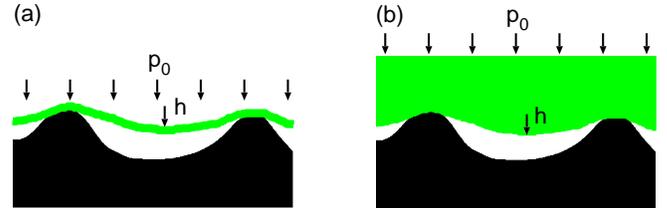}
\caption{
An elastic film (a), and a semi-infinite elastic solid (b), in contact with a randomly rough substrate. The elastic modulus
of the solids in (a) and (b) are such that the displacement $h$ (in response to an external pressure $p_0$)
into the ``cavity'' (with diameter $\lambda$) is the same in
both cases.}
\label{blockcontact}
\end{figure}

If two semi-infinite elastic solids with randomly rough surfaces are squeezed in contact at the (nominal) squeezing pressure $p_0$,
the heat transfer coefficient\cite{PLV}
$$\alpha \approx {p_0 \kappa \over E^* u_0}.\eqno(6)$$
The effective heat transfer coefficient $\kappa$  is defined by
$${1\over \kappa} = {1\over \kappa_0}+{1\over \kappa_1},$$
where $\kappa_0$ and $\kappa_1$ are the heat conductivities of solid ${\bf 0}$ and  solid ${\bf 1}$, respectively.
The effective elastic modulus $E^*$ is defined by
$${1\over E^*} = {1-\nu_0^2\over E_0}+{1-\nu_1^2\over E_1},$$
where $E_0$ and $\nu_0$ are the Young's elastic modulus and the Poisson ratio, respectively, 
for solid ${\bf 0}$, and similar for solid  ${\bf 1}$.
The length parameter $u_0$ in (6) can be calculated 
from the surface roughness power spectrum $C(q)$ as described in Ref. \cite{JCPpers1}.
In general $u_0$ is of order the root-mean-square roughness of the (combined) surface profile.
We have calculated $u_0 \approx 2.5 \ {\rm nm}$ for an amorphous ${\rm SiO_2}$ surface, with the root mean square
roughness $2.5 \ {\rm nm}$ when measured over a surface area $10 {\rm \mu m} \times 10 {\rm \mu m} $. 

Eq. (6) for the heat transfer via the area of real contact is valid
for the case of two semi-infinite elastic solids in non-adhesive contact\cite{PLV}. 
However, the present application involves
an atomically thin elastic film in adhesive contact with a substrate.
It may be possible to extend the theory presented in Ref. \cite{PLV} to this case too, but 
here we will instead present an approximate treatment based on the theory presented above. 

In the derivation of Eq. (6)  it is assumed that the elastic solids
have a thickness much larger than the average separation between the macroasperity contact regions
(see Ref. \cite{PLV}). This condition is not valid for graphene which is only one atom thick.
Nevertheless, using the very simple arguments presented below, 
one can also apply the formula (6) 
for the heat transfer coefficient between
graphene and  ${\rm SiO_2}$. We first note that because of the high (in-plane) heat conductivity of 
graphene ($\kappa \approx 5000 \ {\rm W/mK}$ at room temperature),
the contact resistance will arise on  the ${\rm SiO_2}$-side of the interface, i.e., we can use
$ \kappa =  \kappa_1$ (where $ \kappa_1 \approx 1 \ {\rm W/m K}$ is the heat conductivity of amorphous ${\rm SiO_2}$) in (6).
In the present application there is no external applied pressure but the graphene is bound to the
${\rm SiO_2}$ substrate by adhesion. In a first approximation we can consider the interfacial 
interaction as the sum of the long-ranged van der Waals interaction 
and a short ranged repulsion in the contact regions, where
the electron clouds of the graphene and the substrate overlap. We can apply (6) approximately
to this situation if the pressure $p_0$ is taken as the (average) force per unit aria arising from the
van der Waals interaction. For the separation $d \approx 1 \ {\rm nm}$ one can estimate\cite{Buch} the van der Waals pressure
to be of order $p_0 \approx 10^7 \ {\rm Pa}$. 

Finally, we need the effective elastic modulus $E^*$ to be used in
(6). In the present case we can neglect the deformations of the  ${\rm SiO_2}$ substrate and only include the
deformations of the graphene. We determine $E^*$ as follows:  
Assume first that the surface of a semi-infinite solids (with Young's modulus $E^*$)
is deformed so that it penetrate a distance $h$ into a substrate
cavity with diameter $\lambda$, see Fig. \ref{blockcontact}(b). This cost the elastic energy $\approx E^* \lambda^3 (h/\lambda)^2$.
Here we have used that in a volume element $\sim \lambda^3$ the typical strain is $h/\lambda$.
Let us now instead deform the graphene so it penetrate the same distance $h$ into the cavity, see Fig. \ref{blockcontact}(a). 
This requires stretching the graphene\cite{Carbone}
(i.e., in plane deformation) with $\delta \lambda \approx \lambda (h/\lambda)^2$ over the area $\lambda^2$.
The stored elastic energy is thus $E \lambda^2 t (\delta \lambda / \lambda)^2 \approx E \lambda^2 t  (h/\lambda)^4$,
where $E \approx 10^{12} \ {\rm Pa}$ is the Young's modulus of in-plane deformation of graphite 
and $t$ is the thickness of graphene 
(which is equal to the layer spacing in graphite or $t\approx 0.34 \ {\rm nm}$).
We define $E^*$ so that the two elastic energies are equal which gives 
$$E^* \approx  (t/\lambda ) (h/\lambda)^2 E.\eqno(7)$$
Next note that the energy to deform the film by a distance $h$ can also be written as $\approx p_0 h \lambda^2$, where $p_0$
is the applied pressure. Thus we get $p_0 h \lambda^2 \approx E^* \lambda^3 (h/\lambda)^2$ or $h/\lambda = p/E^*$.
Substituting this in (7) gives
$$E^* \approx (p_0^2Et/\lambda)^{1/3}\eqno(8)$$
Substituting (8) in (6) gives
$$\alpha  \approx \left ( {p_0 \lambda \over E t} \right )^{1/3}{\kappa \over u_0}\eqno(9)$$
This equation shows that $\alpha$ does not depend sensitively on $p_0$ and $\lambda$, which are not accurately known.

Eq. (8) shows that the effective elastic modulus depends on the length scale $\lambda$ which is expected as the
long-range elastic properties of the effective elastic solid and the graphene are different. However, we can estimate
a typical $\lambda$ using $\lambda \approx (\lambda_0 \lambda_1)^{1/2}$, where $\lambda_0$ and $\lambda_1$ are the
wavelength of the longest and shortest surface roughness components. The former is determined by the roll-off wavevector
of the surface roughness power spectrum which is typically $q_0 \approx 3\times 10^7 \ {\rm m}^{-1}$ giving
$\lambda_0 = 2 \pi /q_0 \approx 200 \ {\rm nm}$. The latter is of order a $\lambda_1 \approx 1 \ {\rm nm}$. Thus we get 
$\lambda \approx 14 \ {\rm nm}$. Using (8) we obtain $E^* \approx 2\times 10^8 \ {\rm Pa}$.
Using $p_0 \approx 10^7 \ {\rm Pa}$, $E^*  \approx 10^8 \ {\rm Pa}$, $u_0 \approx 3 \ {\rm nm}$ and $\kappa = 1 \ {\rm W/mK}$
in (6) gives $\alpha \approx 3\times 10^7 \ {\rm W/m^2 K}$ which is very close to the observed value.
In addition, since the thermal conductivity of amorphous ${\rm SiO_2}$ is only weakly dependent on temperature for 
$300 \ {\rm K} < T < 900 \ {\rm K}$ (which result from the short phonon mean free path in the disordered  ${\rm SiO_2}$),
the heat transfer coefficient will be nearly temperature independent in the studied temperature interval, in agreement with
the experimental results of Ref. \cite{Freitag}.
On the other hand the contribution to $\alpha $ from the surrounding gas, 
and from photon tunneling, depends strongly on the temperature.

To summarize, we have studied theoretically the heat transfer between graphene and an amorphous ${\rm SiO_2}$ substrate.
We have found that most of the heat energy flows through the area of real contact, while the heat flow via the surrounding gas,
and from photon tunneling, are both roughly 100 times weaker.

\vskip 0.2cm
{\bf Acknowledgments}

We thank P. Avouris for drawing our attention to Ref. \cite{Freitag}. We thank him and
M. Freitag for information related to the same reference.
B.N.J.P. was supported by Invitation Fellowship Programs for Research in Japan from
Japan Society of Promotion of Science (JSPS).
H.U. was supported  by the Grant-in-Aid for Scientific
Research B (No. 21310086) from JSPS.

\end{document}